\documentclass[usegraphicx,usenatbib]{mn2e}
\usepackage{color,times,float,hyperref}

\newcommand{\Msol}{\ensuremath{M_\odot}}

\newcommand{\etal}{et al.}
\newcommand{\aj}{AJ}
\newcommand{\aanda}{A\&A}
\newcommand{\araa}{ARA\&A}

\newcommand{\apj}{ApJ}

\newcommand{\apjs}{ApJS}

\newcommand{\arxiv}[1]{arXiv:{#1}}
\newcommand{\mnras}{MNRAS}
\newcommand{\newar}{New Ast. Rev.}

\newcommand{\pasp}{PASP}

\newcommand{\prl}{Phys. Rev. Lett.}

\newcommand{\nickel}{{\ensuremath{^{56}\mathrm{Ni}}}}
\newcommand{\cobalt}{{\ensuremath{^{56}\mathrm{Co}}}}
\newcommand{\iron}{{\ensuremath{^{56}\mathrm{Fe}}}}

\newcommand{\kms}{\ensuremath{\mathrm{km~s}^{-1}}}

\newcommand{\MNi}{\ensuremath{M_{\mathrm{Ni}}}}
\newcommand{\Mej}{\ensuremath{M_\mathrm{ej}}}
\newcommand{\Mch}{\ensuremath{M_\mathrm{Ch}}}

\newcommand{\revised}[1]{{\textcolor{black} {#1}}}
\newcommand{\nb}[1]{\ensuremath{^{#1}}}
\newcommand{\code}[1]{{\sc #1}}

\begin{document}


\title[Mass distribution of SN Ia progenitors]
      {The ejected mass distribution of type Ia supernovae: \\
       A significant rate of non-Chandrasekhar-mass progenitors}

\author[Scalzo et al.]
{
    R.~A.~Scalzo\nb{1,2}\thanks{Email: rscalzo@mso.anu.edu.au},
    A.~J.~Ruiter\nb{1,2,3},
    and S.~A.~Sim\nb{2,4}
    \\
    \nb{1} Research School of Astronomy and Astrophysics,
           Australian National University,
           Canberra, ACT 2611, Australia \\
    \nb{2} ARC Centre of Excellence for All-Sky Astrophysics (CAASTRO) \\
    \nb{3} Max-Planck-Institut f\"ur Astrophysik, Karl-Schwarzschild-Str. 1,
           85741 Garching bei M\"unchen, Germany \\
    \nb{4} Astrophysics Research Centre, School of Mathematics and Physics,
           Queen's University Belfast, Belfast, BT7 1NN, UK \\
}

\maketitle

\vspace{-1in}

\begin{abstract}

The ejected mass distribution of type Ia supernovae directly probes progenitor
evolutionary history and explosion mechanisms, with implications for their
use as cosmological probes.  Although the Chandrasekhar mass is a natural mass
scale for the explosion of white dwarfs as type Ia supernovae, models allowing
type Ia supernovae to explode at other masses have attracted much recent
attention.  Using an empirical relation between the
ejected mass and the light curve width, we derive ejected masses \Mej\ and
\nickel\ masses \MNi\ for a sample of 337 type Ia supernovae with redshifts
$z < 0.7$ used in recent cosmological analyses.  We use hierarchical Bayesian
inference to reconstruct the joint \Mej-\MNi\ distribution, accounting for
measurement errors.  The inferred marginal distribution of \Mej\ has a long
tail towards sub-Chandrasekhar masses, but cuts off sharply above 1.4~\Msol.
Our results imply that 25\%--50\% of normal type Ia supernovae are
inconsistent with Chandrasekhar-mass explosions, with almost all of these
being sub-Chandrasekhar-mass; super-Chandrasekhar-mass explosions
make up no more than 1\% of all spectroscopically normal type Ia supernovae.
We interpret the type Ia supernova width-luminosity relation as
an underlying relation between \Mej\ and \MNi, and show that the inferred
relation is not naturally explained by the predictions of any single known
explosion mechanism.
\end{abstract}



\vspace{0.1in}

\section{Introduction}

Type Ia supernovae (SNe~Ia) are widely used as distance indicators in
cosmological studies of the dark energy \citep{riess98,scp99}.
Their accuracy and precision rely on empirical relations between their
luminosity and their light curve width and colour, which suffice to establish
luminosities to 15\% and distances to 7\%
\citep{phillips93,riess96,tripp98,goldhaber01}.
Other standardization methods have also been developed:  reducing sensitivity
to dust using intrinsic colours inferred from optical photometry several weeks
past maximum light \citep{phillips99,conley06,csp10} or from near-infrared
photometry \citep{burns14,mandel09,mandel11}; splitting the sample by
spectroscopic properties \citep{wang09,fk11}; or using spectroscopic
\citep{bongard06,chotard11} or spectrophotometric \citep{bailey09,blondin12}
indicators instead of light curve parameters.

Despite the considerable attention devoted to type Ia supernovae in cosmology,
the evolution of their progenitor systems, their physical state at the time
of explosion, and the explosion mechanisms are not fully understood
\citep[for recent reviews, see][]{wh12,hillebrandt13,pilar14}. 
The long-standing consensus holds that type Ia supernovae must result from
the thermonuclear incineration of at least one carbon-oxygen white dwarf.
This view is supported by several independent lines of evidence:  the layered
composition of SN~Ia ejecta implied by their spectroscopic evolution,
with iron-peak elements in the center and intermediate-mass elements such as
silicon and sulfur on the outside \citep{stehle05,mazzali07}; the range of
star formation histories in SN~Ia host galaxies, showing that SNe~Ia are not
associated exclusively with young stars, as are other types of supernovae
\citep{mannucci06}; and upper limits on the size of the progenitor, based on
upper limits on shock breakout radiation \citep{piro10} shortly after
explosion for the very nearby SN~2011fe \citep{nugent11,bloom12}.
SNe~Ia also must result from interactions of the white dwarf with another
star \citep[e.g.][]{wi73,it84}, since an isolated white dwarf will simply
cool and fade away over billions of years, and that a SN~Ia progenitor is
totally disrupted in the explosion, at least for the spectroscopically
``normal'' SNe~Ia \citep{bfn93,branch06} used in cosmology.

For a long time, the conventional wisdom has favored the delayed detonation
\citep{khokhlov91} of a white dwarf near the Chandrasekhar mass
$\Mch = 1.4~\Msol$ ---
a natural theoretical mass scale at which SNe~Ia might explode --- accreting
and steadily burning material either from a non-degenerate \citep{wi73}
or white dwarf \citep{it84} companion as the explosion mechanism for normal
SNe~Ia \citep{mazzali07}.  However, recent indirect evidence suggests that
this scenario cannot account for all normal SNe~Ia.  Such evidence includes
the lack of observed X-ray emission \citep{gb10} or ionized helium emission
\citep{wg13} associated with nuclear burning of accreted material on the
surface of a white dwarf \citep[see also][]{rds10a,rds10b}, and the
theoretical difficulty of accounting for observed SN~Ia rates using
only Chandrasekhar-mass white dwarfs \citep{vkcj10}.

This pressure has
motivated the study of other SN~Ia channels in which the progenitor
does not need to be very close to \Mch\ for an explosion to take place.
In such models, the ejected mass \Mej\ of a SN~Ia encodes vital information
about the progenitor evolution and explosion mechanism.  They include:
\emph{double detonations}, in which a surface detonation drives a shock
into the interior of a sub-Chandrasekhar-mass white dwarf and ignites it
\citep{ww94,fink10}; \emph{spin-up/spin-down} models, in which angular
momentum of accreted material enables a super-Chandrasekhar-mass white dwarf
to support itself against explosion until internal stresses redistribute
that angular momentum \citep{justham11,rds12}; \emph{violent mergers} of
two white dwarfs in which one or both white dwarfs explode shortly after
the merger event \citep{pakmor12}, generally expected to be
super-Chandrasekhar-mass; and \emph{collisions} of two white dwarfs
\citep{benz89,rosswog09,raskin09} encouraged by Kozai resonances in triple
systems \citep{thompson11,kd12,hamers13}, which may have masses ranging from
sub-Chandrasekhar to super-Chandrasekhar.  Some spectroscopically peculiar,
extremely luminous ($M_B \sim -20$) SNe~Ia
\citep{howell06,hicken07,yamanaka09,tanaka10,scalzo10,taub11,taub13}.
are interpreted as having ejected a super-Chandrasekhar mass of material.
\revised{\citet{mazzali11} find a good fit to the nebular spectrum of
the subluminous, but spectroscopically normal, SN~Ia~2003hv using a
sub-Chandrasekhar-mass model ejecting about 1.0~\Msol; \citet{mh12} make
similar arguments for SN~1991bg.  \citet{mazzali07} suggest a lower limit
of about 1.0~\Msol\ on the mass ejected in a spectroscopically normal SN~Ia.}

In a similar vein, while improved luminosity standardization for cosmology
remains an active area of research, the physical origin of the
width-luminosity (or width-colour-luminosity) relation has yet to be fully
understood.  SNe~Ia are powered by the radioactive decay chain
$\nickel \rightarrow \cobalt \rightarrow \iron$, as has now been
directly confirmed by observation of gamma rays from this decay chain in
the nearby SN~2014J \citep{integral}.  The mass \MNi\ of \nickel\
synthesized in the explosion largely determines the peak bolometric luminosity
\citep[``Arnett's rule'';][]{arnett82} and hence is somehow linked with the
light curve width.  Possible physical drivers include changes in opacity with
temperature \citep{kmh93,hk96} \revised{or with synthesized total iron-peak
element mass \citep{mazzali07}}, variation of the overall
ejected mass \citep{pe00}, or asymmetries arising from hydrodynamic
instabilities in three-dimensional explosion models \citep{wk07,kasen09}.
Improved understanding of the physics behind the width-luminosity relation
would put the measurement of distances to SNe~Ia on sound theoretical
footing, and could uncover new luminosity correlates related to specific
explosion mechanisms or progenitor channels.  Many of the abovementioned
scenarios make predictions for \MNi, or equivalently
the absolute magnitude, of a SN~Ia as a function of progenitor mass.

Recently, \citet{scalzo14a} derived ejected masses for a sample of 19
spectroscopically normal SNe~Ia observed by the Nearby Supernova Factory
(SNfactory) by modeling their bolometric light curves.  They found that
normal SNe~Ia show a range of ejected masses from 0.9--1.4~\Msol,
\revised{roughly compatible with the findings of \citet{mazzali07} and
\citet{mazzali11}}, with about 15\% systematic uncertainty on the absolute mass
scale of the explosion.  They also found a tight correlation between \Mej,
as derived from the bolometric light curve, and the light curve width $x_1$,
as measured from multi-band photometry by the \code{salt2} cosmological light
curve fitter \citep{guy07,guy10}.  These findings confirm the earlier
observational results of \citet{stritz06}, using a similar technique, but
with a factor of three improvement in the precision of the reconstruction.
The method of \citet{scalzo14a} was validated using synthetic data from
contemporary three-dimensional SN~Ia explosion models, demonstrating the
ability to distinguish between explosion scenarios based on the ejected mass.

In this paper, we interpret the SN~Ia width-luminosity relation in terms
of an underlying relation between \Mej\ and \MNi.  We apply the
\citet{scalzo14a} \Mej-$x_1$ relation and a version of Arnett's rule to
estimate \Mej\ and \MNi\ for a
large sample of SNe~Ia with redshift less than 0.7.
We then use hierarchical Bayesian inference to reconstruct the \emph{joint}
\Mej-\MNi\ \emph{distribution} for this sample.  This establishes a proof
of principle for the direct comparison of observations with predicted
theoretical distributions, based on our best contemporary binary population
synthesis and numerical explosion models.  Finally, we derive a relative rate
of non-Chandrasekhar-mass SNe~Ia based on this distribution and discuss the
implications for the progenitor evolution and explosion mechanism(s).


\section{Hierarchical Bayesian inference of the joint \Mej-\MNi\ distribution}
\label{sec:bayes}

To derive an accurate \Mej-\MNi\ joint distribution, we require a pure,
unbiased sample of spectroscopically normal SNe~Ia.  Numerous observational
selection effects prevent any real sample from fully realizing this ideal.
For our purposes, we require that the supernovae we use are spectroscopically
confirmed \emph{normal} SNe~Ia that could appear in a cosmology analysis.
We also require supernovae that have been discovered in a search not targeting
specific host galaxies, to avoid environmental biases on the distribution of
intrinsic SN~Ia properties.  The recent Joint Light Curve Analysis sample
\citep[JLA;][]{betoule14}, drawing from the untargeted
SDSS \citep{sdss} and SNLS \citep{snls} searches,
provides a good initial source of spectroscopically confirmed SNe~Ia that
have in fact been used in a cosmology analysis.  We confine our attention to
volume-limited subsamples of the \citet{betoule14} SNe~Ia to avoid Malmquist
bias on the light curve width and colour, and we include only SNe~Ia that pass
all of the \citet{betoule14} data quality cuts and have reliable \code{SALT2}
light curve fits.  We select SDSS SNe with $z < 0.2$ (197~SNe) and SNLS SNe
with $z < 0.7$ (140~SNe), for a total of 337 SNe~Ia.

The JLA SNe~Ia do not, in general, have sufficiently broad wavelength coverage
in multi-band photometry to construct full bolometric light curves; while
corrections can be made for missing ultraviolet and near-infrared flux,
measurements equivalent to rest-frame $UBVRI$ are needed.  They are also,
in general, not observed out to the \cobalt-dominated phase needed to apply
the full mass reconstruction method of \citet{scalzo14a}.  However, the
correlation between \Mej\ and \code{SALT2} $x_1$ found by \citet{scalzo14a}
can be used to estimate \Mej\ without full bolometric light curves.
The \citet{scalzo14a} SNe~Ia also show a strong correlation between inferred
\MNi\ and absolute magnitude $M_B$, as expected from Arnett's rule.
We can therefore use such relations to transform the \code{SALT2} light curve
fit parameters $(m_B, x_1, c)$ directly into intrinsic properties
$(\Mej, \MNi)$ within the JLA best-fitting cosmology.

To find the transformation equations, we perform least-squares fits using
\Mej\ and \MNi\ values from run F in table 6 of \citet{scalzo14a}, including
all spectroscopically normal SNe~Ia and incorporating measurement errors in
both the dependent and independent variables for each fit.  We obtain
\begin{equation}
\Mej/\Msol = (1.322 \pm 0.022) + (0.185 \pm 0.018) \, x_1
\label{eqn:mejvx1}
\end{equation}
with $\chi^2/\nu = 13.3/16 = 0.829$ and 6\% RMS dispersion, and
\begin{equation}
\log \MNi/\Msol = -0.4 (M_B + 19.841 \pm 0.020)
\label{eqn:mnivmb}
\end{equation}
with $\chi^2/\nu = 3.794/16 = 0.237$ and 8\% RMS dispersion.  Here $M_B$ has
been corrected for host galaxy dust extinction assuming a \citet{cardelli}
extinction law with $R_V = 3.1$.  The value of \MNi\ from the full bolometric
light curve fit can be even more accurately predicted (to 4\% RMS) by adding
to Equation~\ref{eqn:mnivmb} a term linear in extinction-corrected colour,
i.e., a bolometric correction at maximum light.  Such a correction
requires an independent estimate of the dust extinction for each SN, which
is problematic for the JLA sample.
We can proceed using Equation~\ref{eqn:mnivmb} assuming a scenario similar
to that studied by \citet{scolnic13}, in which colour variation among SNe~Ia
is due almost entirely to dust with $R_V = 3.1$, and neglecting the bolometric
correction.  While by no means absolutely certain, there is some independent
evidence that such a scenario may not be far from the truth
\citep[e.g.,][]{chotard11}.

Multilinear regressions against all three \code{SALT2} light curve variables
do not significantly improve predictions for \Mej\ and \MNi\ over
these single-parameter relations.  In other words, \Mej\ is most directly
related to $x_1$, \MNi\ is most directly related to $M_B$
(after correction for extinction), and any other correlations, such as the
width-luminosity relation itself, are secondary to these, at the
level of detail of the \citet{scalzo14a} modeling.  In general, the dominant
uncertainties on \Mej\ and \MNi\ are systematic, and are associated with
limitations in the bolometric light curve modeling rather than observed
scatter around these relations.  The unknown functional
form of the ejecta density profile is the largest driver of uncertainty in
\Mej; if it is the same for all SNe~Ia, its first-order influence is to
change the zeropoint of Equation~\ref{eqn:mejvx1}, without affecting the
slope or statistical significance.  The largest uncertainty in \MNi\ is
the extent of radiation trapping near bolometric maximum light.  This is
represented by a factor $\alpha$ representing the ratio of the bolometric
luminosity to the rate of energy input from radioactive decay, which may vary
from SN to SN, but is believed to be close to 1 for a variety of explosion
models \citep{arnett82,kmh93,nugent95,hk96,howell09,blondin13}.

Simply applying Equations~\ref{eqn:mejvx1} and \ref{eqn:mnivmb} to determine
(\Mej, \MNi) from $(m_B, x_1, c)$ for each SN, and propagating Gaussian
uncertainties, produces an \emph{observed} two-dimensional
distribution that has been distorted by measurement errors.  To infer the
\emph{error-free} distribution, we use the forward-modeling algorithm of
\citet{hogg10}.  This is an importance-sampling method that requires samples
from the $(\Mej, \MNi)$ joint probability distribution for
each SN~Ia.  We first generate $(M_B, x_1, c)$ samples using
(the Cholesky decomposition of) the covariance matrix of the \code{SALT2}
light curve fit for each SN as given in \citet{betoule14}.  Inspired by the
BALT prescription of \citet{scolnic13}, we interpret variation in $c$ as
due entirely to dust extinction with $R_V = 3.1$, with
\begin{equation}
P(c) = e^{-(c-\bar{c})^2/\tau_S^2}
\end{equation}
for $c > \bar{c}$, with $\bar{c} = -0.1$ and $\tau_S = 0.11$.  We impose
this prior by random rejection of samples as they are generated.
After applying the extinction correction to $M_B$ for each sample, we use
Equations \ref{eqn:mejvx1} and \ref{eqn:mnivmb} to convert $(x_1, M_B)$
directly to ($\Mej, \MNi$).  We apply an additional 6\% random Gaussian
scatter to \Mej\ after sampling, to represent dispersion of the results from
the full bolometric light curve fit around Equation \ref{eqn:mejvx1}.
We also apply an additional 20\% random Gaussian scatter to \MNi,
to represent potential variation in the radiation trapping factor
$\alpha$ between different SNe~Ia, typical of the priors used in
\cite{scalzo14a}.  Equation~\ref{eqn:mnivmb} assumes
$\alpha = 1$, and while $M_B$ is a good predictor of bolometric luminosity
for SNe~Ia, the true spread in \MNi\ may be larger than what we infer.

We parametrize the intrinsic $(\Mej, \MNi)$ distribution as a sum of
Gaussians, centered on a grid
with $0.7 < \Mej/\Msol < 1.7$ (spacing 0.05) and $0.3 < \MNi/\Msol < 0.8$
(spacing 0.1), with width in each direction equal to the grid spacing,
resulting in 100 parameters (the ``bin'' heights).  For each SN~Ia, we draw
$k = 200$ samples $(\Mej, \MNi)$ based on the \code{SALT2} light curve fit,
and use these to calculate the \citet{hogg10} likelihood for the parameters
describing the intrinsic distribution.  We minimize this likelihood to provide
a plausible initial guess for the form of the intrinsic distribution, then
sample the full joint distribution of the parameters using the
affine-invariant Markov chain Monte Carlo code \code{emcee} \citep{emcee}.

Once we have inferred the intrinsic distribution $P(\Mej, \MNi)$, we can
re-apply it as a prior to find new, hierarchical estimates for
\Mej\ and \MNi\ for each SN~Ia.  The bolometric light curve fits in
\citet{scalzo14a} incorporated knowledge of SN~Ia
explosion physics into the likelihood $P(\mathrm{data}|\Mej, \MNi)$,
including constraints between \Mej, \MNi, and other nuisance parameters
such as the kinetic energy of the explosion; however,
they used an uninformative (uniform) prior $P_0(\Mej, \MNi)$.
We can thus approximate
\begin{equation}
P(\Mej, \MNi | \mathrm{data})
   = \frac{P(\mathrm{data} | \Mej, \MNi) \, P(\Mej, \MNi)}{P(\mathrm{data})}
\end{equation}
by re-weighting the $(\Mej, \MNi)$ samples for each SN~Ia by the inferred
$P(\Mej, \MNi)$.  While single-point estimates use only information about the
light curve, the hierarchical estimate incorporates information about where
the other SNe~Ia in the sample are found in the $(\Mej, \MNi)$ plane.
The distribution of hierarchical estimates is more concentrated than the
inferred error-free distribution because of the measurement uncertainties in
the individual points; in the limit of large errors, the posterior probability
for each SN simply becomes the prior.

\begin{figure*}
\resizebox{\textwidth}{!}{\includegraphics{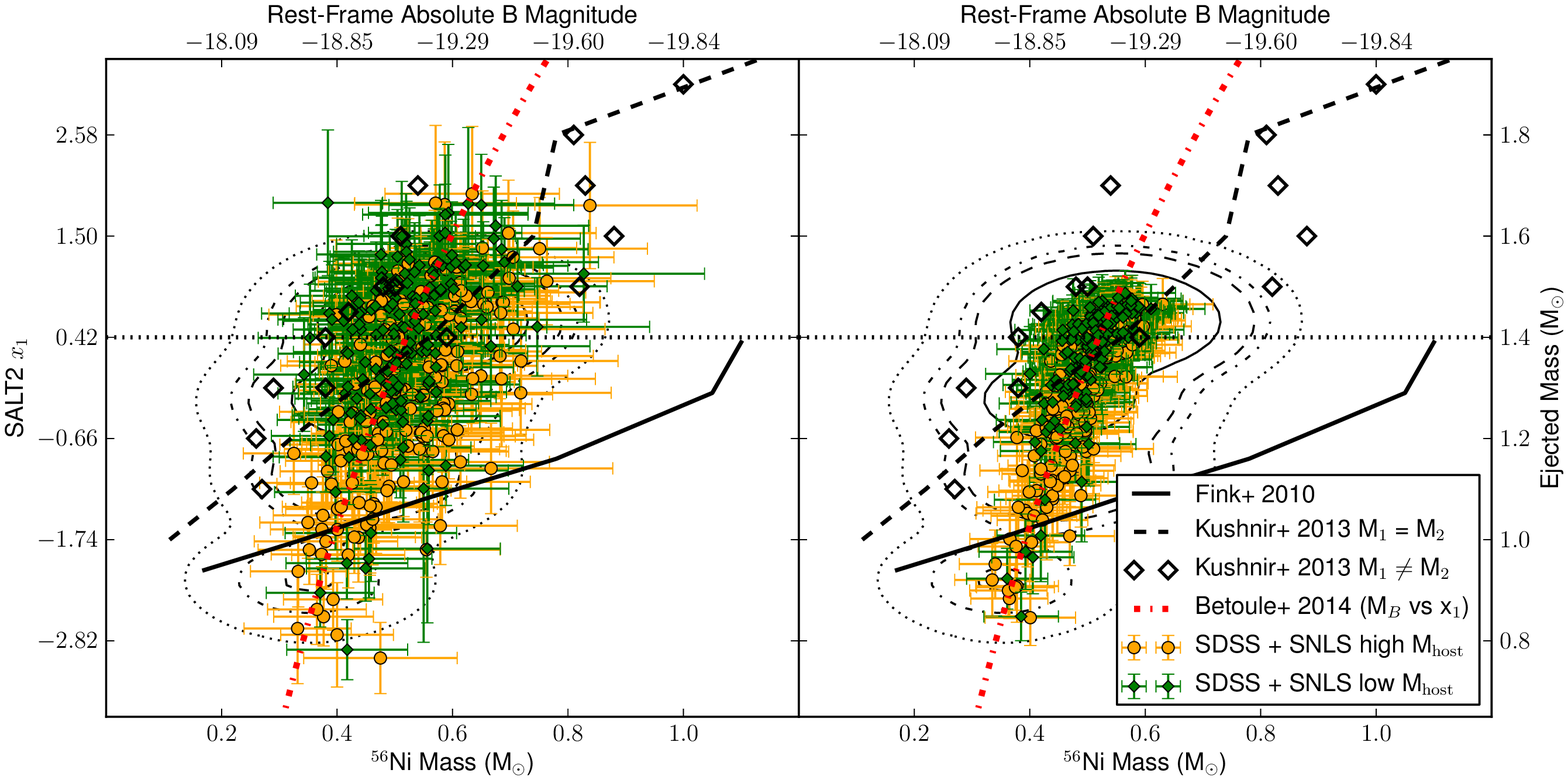}}
\caption{\small Joint \Mej-\MNi\ distribution for 337 SNe~Ia.
Black open diamonds:  non-equal-mass white dwarf collisions of
\citet{kushnir13}.  Dashed line:  equal-mass white dwarf collisions of
\citet{kushnir13}.  Solid line:  double detonations of \citet{fink10}.
Contours:  68\%, 90\%, 95\%, and 99\% confidence regions, bounded by
level sets of probability density, for the \Mej-\MNi\ distribution.
Symbols with error bars:  Bayesian mass estimates for individual SNe~Ia
(green: $M_\mathrm{host} < 10^{10}$~\Msol;
orange: $M_\mathrm{host} > 10^{10}$~\Msol).
Left:  estimates of \Mej\ and \MNi\ using Equations \ref{eqn:mejvx1}
and \ref{eqn:mnivmb} with an uninformative prior $P_0(\Mej, \MNi)$;
right:  imposing as a prior the inferred joint \Mej-\MNi\ distribution
$P(\Mej, \MNi)$ shown by the contours (right).}
\label{fig:mejvmni}
\end{figure*}

Figure~\ref{fig:mejvmni} shows joint confidence regions in the (\Mej, \MNi)
plane for the combined sample, bounded by level sets of probability density,
together with the best estimate of these parameters (and uncertainties)
for each supernova.  The predictions of two contemporary explosion models
are also shown:  double detonations \citep{fink10} and white dwarf collisions
\citep{kushnir13}.  The red dot-dashed line shows the transformation into
the \Mej-\MNi\ plane of the best-fitting SN~Ia $M_B$-$x_1$ relation from
\citet{betoule14}, which the hierarchical estimates for each SN~Ia trace
with about 8\% dispersion in \MNi; this is of a similar order to the
dispersion found by \citet{scolnic13} under similar assumptions about SN~Ia
intrinsic colour and extinction.  At this level of detail,
the relation still appears to be a single-parameter family, with no clear
evidence for multiple sub-populations with different explosion properties
(e.g. different intrinsic luminosities).

\begin{figure}
\resizebox{0.48\textwidth}{!}{\includegraphics{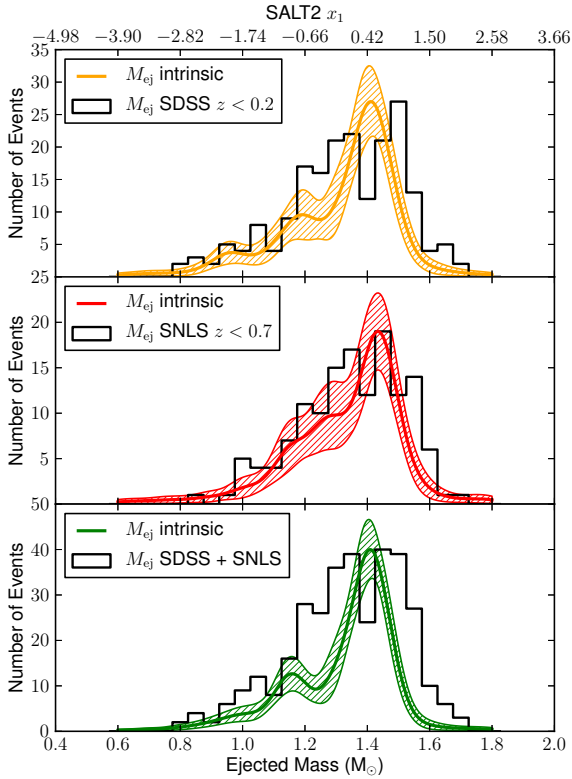}}
\caption{\small Marginal distribution of \Mej\ over subsets of SNe~Ia used in
the \citet{betoule14} cosmological analysis.  Top:  SDSS SNe~Ia with $z < 0.2$.
Middle:  SNLS SNe~Ia with $z < 0.7$.  Bottom:  Union of the two subsets.
Histograms show the original distribution of mean values as inferred directly
from the \code{SALT2} light curve parameters.  Coloured curves with hatched
bands show the mean and 68\% CL variation of the intrinsic distribution,
parametrized as a sum of Gaussians.}
\label{fig:mejhist}
\end{figure}

For a simpler view, Figure~\ref{fig:mejhist} shows the marginal distributions
of \Mej\ inferred separately from the SDSS, SNLS, and combined samples.
The marginal \Mej\ distributions show that the distribution of $x_1$ for
SDSS supernovae is similar to that for SNLS supernovae.  Both distributions
show a tail towards lower \Mej, and both show a sharp cutoff of the
distribution above about 1.45~\Msol.  The peak of the marginalized \Mej\
distribution is close to 1.4~\Msol.  The hint of structure at around
1.2~\Msol\ is tantalizing, but not highly statistically significant;
we discuss the possible implications in \S\ref{sec:interpretation} below.

We caution that these results rely on the accuracy of the assumptions of
\citet{scalzo14a}, from which Equations \ref{eqn:mejvx1} and \ref{eqn:mnivmb}
were derived:  spherically symmetric ejecta, with a stratified composition
and a universal functional form for the radial density profile, for all
normal SNe~Ia.  While these
properties agree with conventional wisdom for the modeling of SN~Ia
explosions and can be supported observationally, there may be variation
among real SNe~Ia that is not captured by our model.  \citet{scalzo14a}
had some difficulty accurately reproducing \MNi\ for some highly asymmetric
SN~Ia explosion models, although inferred \MNi\ tracks modeled \MNi\ for
angle-averaged light curves.  Similarly, if the ejecta density profiles of
SNe~Ia vary systematically with light curve width, this could affect the
slope of Equation~\ref{eqn:mejvx1} and hence the shapes of our inferred
distributions.

\section{Relative rates of non-Chandrasekhar-mass SNe~Ia}

The calibration of the zeropoint of Equation~\ref{eqn:mejvx1} is also subject
to uncertainty at the 10--15\% level \citep{scalzo14a}.  In this context,
the sharp peak in our inferred distribution of \Mej\ at 1.4~\Msol\ carries
weight in any physical interpretation of the distribution.  \citet{scalzo14a}
present strong evidence that SNe~Ia do not all explode at \Mch.  However,
if the calibration of our ejected mass scale is very far from what we assume,
we then have the additional challenge of explaining why most SNe~Ia explode
at a preferred mass \emph{other} than \Mch.  Some theoretical scenarios,
such as the double-degenerate violent merger scenario, may provide 
motivation for such a peak; we discuss these cases in
\S\ref{sec:interpretation}.

\begin{figure}
\label{fig:mejvx1}
\resizebox{0.5\textwidth}{!}{\includegraphics{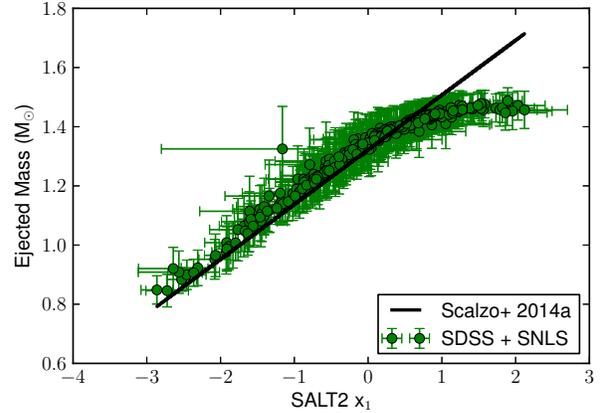}}
\caption{\small The \Mej-$x_1$ relation for our sample.  Symbols with error
bars are hierarchical Bayesian estimates for individual SNe~Ia as shown in
Figure~\ref{fig:mejvmni}.  Solid line:  Equation~\ref{eqn:mejvx1}.}
\end{figure}

Figure~\ref{fig:mejvx1} shows how the functional dependence of \Mej\ on $x_1$
changes when the prior $P(\Mej, \MNi)$ is taken into account.
We see that the \citet{scalzo14a} \Mej-$x_1$ relation changes little
for SNe with $x_1 < +0.5$, although it underestimates the mass
slightly relative to the hierarchical estimate, since faster-declining SNe~Ia
are intrinsically less numerous.  For slower-declining supernovae,
the relation begins to flatten until it is consistent with a constant
($\Mej = 1.45~\Msol$) for all SNe~Ia with $x_1 > +1$.  The spectroscopically
peculiar 1991T-like SNe~Ia are usually found in this range:
SN~1991T \citep{filippenko92,phillips92},
SN~2003fg \citep[SNLS-03D3bb;][]{howell06},
SN~2007if \citep{scalzo10,yuan10}, LSQ12gdj \citep{scalzo14b}, and three of
the five additional super-Chandrasekhar-mass SN~Ia candidates considered in
\citet{scalzo12} have $x_1 > +1$, and among these only SN~2003fg and SN~2007if
have been established as having super-Chandrasekhar-mass ejecta at high
confidence.  Since the 1991T-like sub-classification becomes less secure
in post-maximum spectra \citet{li11}, it is possible that some of these
very-slowly-declining SNe~Ia are actually 1991T-like.


We can make a very conservative estimate of the relative rate of non-\Mch\
explosions by considering only \emph{individual} SNe~Ia which are incompatible
with a Chandrasekhar-mass explosion,
based on the hierarchical Bayesian estimates of \Mej\ and \MNi.  We estimate
errors in the relative rates by drawing samples $P_j(\Mej, \MNi)$ from the
posterior for the inferred $P(\Mej, \MNi)$, and using these to make different
realizations of the hierarchical estimates for each SN.  Out of 337 SNe~Ia,
we find $93.6 \pm 13.2$, or $(27.8 \pm 3.9)$\%, have $\Mej < \Mch$ at 95\% CL
or greater; in contrast, only $4.5 \pm 1.4$, or $(1.3 \pm 0.4)$\%, have
$\Mej > \Mch$ at 95\% CL or greater.  The boundary between Chandrasekhar-mass
and sub-Chandrasekhar-mass systems is thus two standard deviations from
1.4~\Msol, or about 1.23~\Msol\ assuming 6\% dispersion around
Equation~\ref{eqn:mejvx1}.

For a more realistic estimate of the \emph{total} number of
non-Chandrasekhar-mass SNe~Ia, we set plausible boundaries for the
Chandrasekhar-mass regime based on explosion physics, and then count the
number of SNe~Ia lying outside this region; individual SNe~Ia can scatter
either into this region or out of it, based on measurement errors.
With regard to the lower mass bound,
\citet{lesaffre06} simulate the evolution of accreting white dwarfs through
the carbon flash, and find ignition takes place at central densities above
$2 \times 10^9$~g~cm$^{-3}$.  In three-dimensional simulations of
Chandrasekhar-mass delayed detonations, \citet{seitenzahl11} and
\citet{krueger12} consider central densities as low as $10^9$~g~cm$^{-3}$,
corresponding to a mass of 1.36~\Msol for a non-rotating white dwarf.
On the high-mass end, white dwarfs in rigid rotation could
be as massive as 1.5~\Msol\ \citep{anand65,roxburgh65}; white dwarfs more
massive than this would correspond to the super-Chandrasekhar-mass,
differentially rotating white dwarfs of \citet{yl05}.

We therefore take $1.35 < \Mej/\Msol < 1.5~\Msol$ to be
``Chandrasekhar-mass'' --- massive enough to ignite spontaneously,
but still close to the rigidly rotating regime.
These boundaries yield $174.1 \pm 15.2$
Chandrasekhar-mass, $159.8 \pm 15.1$ sub-Chandrasekhar-mass,
and $3.1 \pm 0.9$ super-Chandrasekhar-mass events,
or a sub-Chandrasekhar-mass rate of ($47.4 \pm 4.5$)\%.


Both of these relative rate estimates are consistent with the expectation
that sub-Chandrasekhar-mass SNe~Ia are relatively common \citep{scalzo14a},
while super-Chandrasekhar-mass SNe~Ia are quite rare \citep{scalzo12}.
Thus, if the peak of our derived \Mej\ distribution actually occurs at \Mch,
25\%--50\% of all normal SNe~Ia eject sub-Chandrasekhar masses.
This number can increase if our absolute mass scale is miscalibrated
and the peak in the distribution represents a lower mass.



\section{Interpreting the joint \Mej-\MNi\ distribution}
\label{sec:interpretation}


The identification of ejected mass as the primary factor determining light
curve shape provides a fresh interpretation of previous work on SN~Ia
progenitors and their evolution over cosmic time, a few examples of which
we give here.  SN~Ia rates \citep{sb05} and delay time distributions
\citep{mannucci06} are explained well by models with two populations of
SNe~Ia, split by host galaxy properties such as stellar mass and star
formation rate \citep{sullivan06} and metallicity \citep{howell09}.
\citet{howell09} found that fast-declining
(\code{SALT1} $s < 0.9$, \code{SALT2} $x_1 < -0.9$) SNe~Ia happen almost
exclusively in high-metallicity galaxies ($12 + \log(\mathrm{O/H}) > 8.8$);
these correspond to SNe~Ia with $\Mej < 1.15~\Msol$, and we see a related
preference of these SNe~Ia for high-mass (metal-rich, old, passive)
galaxies in our Figure~\ref{fig:mejvmni}.
\citet{howell07} investigated the possible evolution of the
light curve width with redshift; figure~2 of that work maps directly on to
our Figure~\ref{fig:mejhist}, although we find a less pronounced evolution
trend.  The $z < 0.1$ sample of \citet{howell07} was drawn from targeted
searches that sampled different host galaxy environments from the untargeted
SDSS and SNLS searches, which may be enough to explain the discrepancy.

\citet{piro14} considered different possible theoretical forms of an
underlying \Mej-\MNi\ relation, motivated by the same explosion models we
consider here.  However, \citet{piro14} compared to data by using the
width-luminosity relation to transform light curve width $\Delta m_{15}$
to \MNi\ \citep{mazzali07}, then transforming \MNi\ to \Mej\ assuming that
all SNe~Ia come from a single explosion mechanism.  Our work goes beyond
this by providing more direct
estimates of both \Mej\ and \MNi, and by estimating the actual relative rate
of SNe~Ia inconsistent with Chandrasekhar-mass delayed detonations
(which thus \emph{require} an explanation by some alternative scenario).
Our framework also enables comparison to explosion scenarios that predict
only a \Mej-\MNi\ joint distribution rather than a one-to-one relation;
this may be useful when including uncertainties in \nickel\
production mentioned by \citet{piro14},
e.g., impact parameters in white dwarf collisions.

In this section we consider the specific implications of our findings for a
range of explosion scenarios.  Different uncertainties in our analysis of
$P(\Mej, \MNi)$ have different impacts on particular explosion models,
which we include in our discussion.


\subsection{Chandrasekhar-mass delayed detonations}

A large relative rate of Chandrasekhar-mass SNe~Ia would readily explain
the peak we observe in the \Mej\ distribution near 1.4~\Msol.  In fact,
conventional Chandrasekhar-mass scenarios are the \emph{only} known progenitor
scenario that naturally result in such a peak, providing a strong motivation
to believe that they contribute significantly to the overall SN~Ia rate.
In addition, \citet{seitenzahl13b} showed that a large fraction
($\sim 50$\%) of SNe~Ia must explode at or near \Mch\ in order to explain
the solar abundance of manganese observed in the Galaxy.  The best-studied
explosion mechanism in this case is a delayed detonation,
taking place within the single-degenerate scenario.

In the single-degenerate scenario, a white dwarf accretes from a disc fed
by Roche lobe overflow from a binary companion \citep{wi73}.  As it accretes
mass, the white dwarf primary must therefore also accrete angular momentum.
This will tend to increase its rotation rate and support it 
against collapse or explosion.  Thus the Chandrasekhar-mass scenario could,
with little modification, account for a modest range of \Mej,
e.g. 1.4--1.5~\Msol\ for solid-body rotation \citep{anand65,roxburgh65}.
There are arguments that differentially rotating white dwarfs should not
occur in nature \citep[e.g.,][]{piro08}, but if it does occur, similar
explosion scenarios could be realized for higher, super-Chandrasekhar masses
\citep{justham11,hachisu11,rds12}.
Within this picture, \nickel\ production would have to be controlled by some
other parameter, to account for the $\sim 0.2$~\Msol\ spread in \MNi\ for
systems with ejecta mass near \Mch.  This can be fairly easily achieved in
model sequences that vary properties of the ignition or the transition to
detonation \citep{kasen09,seitenzahl13a,sim13}.

However, Chandrasekhar-mass explosions cannot explain all SNe~Ia if \Mej\
is linked directly to light curve shape.  \citet{scalzo14a} discuss the
limitations of the original bolometric light curve modeling on which Equations
\ref{eqn:mejvx1} and \ref{eqn:mnivmb} is based.  The challenge
\revised{in reproducing a normal SN~Ia with a fast-declining
(\code{SALT2} $x_1 < -1$) light curve with a Chandrasekhar-mass model is to
ensure that the decline from maximum to late times
($> 60$~days after explosion) is also
adequately reproduced.} Reducing the Compton depth by shifting \nickel\ to
higher velocities, e.g., by displacing \nickel\ with stable iron as a result
of neutronization \citep{hoflich04,motohara06}, is not sufficient to do this;
a substantial fraction of \nickel\ at high velocities would be needed.

Alternatively or in addition, the radiation-trapping factor $\alpha$ could be
much greater than 1; reproducing the \citet{scalzo14a} bolometric light curve
of SN~2008ec, for instance, would require $\alpha > 1.5$.
For centrally concentrated \nickel\ and constant opacity, $\alpha = 1$,
and shifting \nickel\ outwards tends to reduce $\alpha$ \citep{pe00}.
In the context of 1-D models, the same factors that increase $\alpha$ also
tend to increase the diffusion time.  A rapid release of trapped energy could
in principle be achieved by a dramatic drop in the opacity before or near
maximum light, perhaps driven by cooling of the ejecta \citep{kmh93},
but a large effect is needed.  \citet{blondin13}, using the
detailed radiation-transfer code \code{cmfgen}, in fact find little
dependence of $\alpha$ on \MNi\ for a sequence of 1-D Chandrasekhar-mass
delayed detonations; they also find \revised{broader bolometric light curves
for SNe~Ia with less \nickel}, in contrast to what \citet{scalzo14a} observe
\revised{(see figure~6 of the latter paper)}.
Finally, highly aspherical ejecta can produce large variations in the
maximum-light bolometric luminosity depending on the viewing angle, but
these effects occur more often in non-Chandrasekhar-mass models such as
violent mergers \citep{pakmor12,moll14}, and the low continuum polarization
of most SNe~Ia argues against highly aspherical explosions \citep{ww08}.

Explosions at or near the Chandrasekhar mass limit might also result from
white dwarfs fed by accretion of material from a disrupted white dwarf
secondary, resulting from a massive double-degenerate merger event
\citep{it84}.  In this case, the ejected mass may exceed \Mch\ when the
mass of the secondary is included.  However, it may be more likely that the
final result of such a merger is collapse to a neutron star, rather than
explosion \citep{nk91}.  Moreover, simulations of supernovae surrounded by
dense carbon-oxygen envelopes, as might arise from such mergers, result in
explosions that may not resemble normal SNe~Ia \citep{fryer10,bs10}.


\subsection{Chandrasekhar-mass pure deflagrations}

Pure deflagrations \revised{present a mechanism to produce weak explosions
of Chandrasekhar-mass progenitors, possibly leaving a bound remnant
\citep{livne05}.}  Simulations of pure deflagrations in Chandrasekhar-mass
white dwarfs have recently been carried out by
\citet{jordan12} 
and by \citet{fink14}. 
These simulated explosions predict a mean trend in which \Mej\
varies significantly with moderate change in \MNi;
for example, Table~1 of \citet{fink14} shows models with
$0.86 < \Mej/\Msol < 1.4$ and $0.26 < \MNi/\Msol < 0.38$.  However,
deflagrations with bound remnants produce too little \nickel\ to explain
observations of normal SNe~Ia.  Additionally, synthetic observables
(light curves and spectra) for \revised{these models} have been shown
to give a poor match to normal SNe~Ia; instead they match fairly well to
observations of the members of the peculiar class of 2002cx-like SNe~Ia
\citep{jha06,phillips07,kromer13a,long13}.
\revised{\citet{sahu08} find that observations of the 2002cx-like SN~Ia~2005hk
is consistent with a Chandrasekhar-mass pure deflagration in which the
white dwarf is completely disrupted, leaving no remnant.}

\revised{In any event, pure deflagrations are at present being invoked to
explain spectroscopically peculiar, underluminous SNe~Ia with low explosion
energy, rather than spectroscopically normal SNe~Ia appearing on the Hubble
diagram.} We therefore consider it unlikely
that Chandrasekhar-mass pure deflagrations have an important role in
explaining the joint \Mej-\MNi\ distribution in Figure~\ref{fig:mejvmni}.


\subsection{Sub-Chandrasekhar-mass double detonations}

Detonations of sub-Chandrasekhar-mass white dwarfs can produce bright
explosions (high \MNi) for relatively low \Mej.  A detonation of a
hydrostatic carbon-oxygen white dwarf with mass in the range 1.0--1.15~\Msol\
will completely unbind the white dwarf and yield \nickel\ mass in a range
that brackets the values inferred for our sample of objects \citep{sim10}.
Synthetic light curves and spectra for such explosions provide a reasonable
match to the properties of normal SNe~Ia \revised{near maximum light}
\citep{sim10}, adding weight to the possibility that sub-Chandrasekhar-mass
detonations might produce normal SNe~Ia in nature.

The best known sub-Chandrasekhar-mass explosion scenario is the
\emph{double-detonation} scenario, in which detonation of a helium layer on
the surface of a carbon-oxygen white dwarf drives a compression shock into
the white dwarf's interior, eventually causing it to detonate in turn
\citep{ww94,fink10,wk11}.
The necessary helium could be accreted slowly from a non-degenerate
\citep{it91} or degenerate \citep{ruiter14} companion, or could arise from
a merger with a helium white
dwarf \citep{pakmor13}.  The locus of points in Figure~\ref{fig:mejvmni}
with \Mej\ in the range 0.9--1.1~\Msol\ are consistent with the prediction
of the double-detonation model of \citet{fink10}.  The apparent excess of
events near this mass in Figure~\ref{fig:mejhist} could also arise from
this channel, although future analyses with larger samples will clarify
whether this excess is real.

However, the double-detonation scenario struggles to match the whole
population or explain the joint distribution.  A generic prediction of
detonations in cold, hydrostatic white dwarfs is that \MNi\ should be
very sensitive to the progenitor mass, producing a steep dependence of \MNi\
on \Mej.  In contrast, our analysis suggests that \MNi\ varies relatively
weakly as a function of the ejected mass for SNe~Ia in nature.
\citet{piro14} make the similar point that only double detonations occurring
in a very limited range of \Mej\ produce \MNi\ consistent with normal SNe~Ia.
They suggest that the \nickel\ production in a double-detonation explosion
may be sensitive to other factors depending on the dynamical details of the
accretion or merger process \citep[e.g.,][]{zhu13,dan14}.

It is also conceivable, but unlikely, that the main peak in our inferred mass
distribution actually occurs at a lower mass, due to modeling uncertainties
as mentioned in \S\ref{sec:bayes}.  The ejecta density profiles for the
\revised{faintest, fastest-declining SNe~Ia in our distribution} would have
to be even more centrally peaked than an exponential to increase the Compton
depth.  This would bring our assumptions into tension with widths of nebular
lines observed in SNe~Ia \citep[$\sim 10^4$~\kms;][]{mazzali98}.


\subsection{Violent white dwarf mergers}

\citet{pakmor10,pakmor11} considered ``violent'' mergers of two carbon-oxygen
white dwarfs, in which the explosion happens promptly during the dynamical
merger process, as progenitors of peculiar, subluminous SNe~Ia.  Later,
\citet{pakmor12} presented a new violent merger model with $\MNi = 0.6~\Msol$,
capable of representing a normal SN~Ia.  The findings of these simulations
suggested that the mass of the primary white dwarf was the most important
parameter determining \nickel\ production, since the less massive secondary is
totally disrupted in the explosion and its remnants do not achieve high enough
densities to burn to the iron peak.  Operating on this assumption,
\citet{ruiter13} used binary population synthesis models to predict the
distribution of \MNi\ for violent mergers, comparing to the observed absolute
magnitude distribution from \citet{li11}.  They found the predicted
distribution to be more or less peaked around $\MNi \sim 0.6~\Msol$
($\Mej \sim 1.1~\Msol$), depending on assumptions about the mass ratio
necessary to ignite a detonation in the primary white dwarf.

The physical explosion mechanism in the \citet{pakmor12} violent mergers is
similar to that in a double detonation, and the parameter space is
approximately bounded from below by the \citet{fink10} curve in our
Figure~\ref{fig:mejvmni}.  However, \Mej\ could vary significantly from
system to system, depending on the fate of the secondary.  If the secondary
is always completely disrupted
\citep[as in][]{pakmor10,pakmor11,pakmor12}, then \Mej\ will be
significantly higher for given \MNi\ than in double detonation models,
but may still vary from system to system because of differences in the
secondary mass.  Even more variation in the total mass could occur if only
a portion of the secondary mass is unbound, which may be possible in some
cases.  The conclusion of \citet{scalzo14a} that violent mergers could not
explain most normal SNe~Ia was based on the assumption that a mass ratio
greater than 0.8 was necessary for ignition, and that the entire mass of the
system was ejected in the explosion, as in \citet{pakmor12}.
However, no comprehensive parameter studies for the fate of the
secondary star in this scenario have yet been made.  In particular, much of
the material from the disrupted secondary star may lie at low velocities
\citep{pakmor12,roepke12,kromer13b}, in tension with late-time observations
of normal SNe~Ia \citep[but see][]{kromer13a}.

In a separate study, \citet{moll14} performed three-dimensional simulations
of three different violent mergers.  In these models, \MNi\ also varies with
\Mej, though with a different functional dependence than that predicted
by \citet{ruiter13}; the ejecta are highly aspherical, even toroidal, and the
light curve varies strongly with viewing angle.  The effective $\alpha$,
relative to spherical models such as \citet{arnett82}, lie in the range
0.6--1.7, clearly violating Arnett's rule; the 1.95-\Msol\ merger model of
\citet{pakmor12} showed similar variation between different lines of sight.
\revised{Similar} variation could in principle produce the narrow light curves 
\revised{seen by \citet{scalzo14a} for the SNe~Ia they infer to be the least
massive}, particularly SN~2005el, which presented the most tension with
expectations from spherical models.  However, all of the \citet{moll14}
mergers are brighter at late times than \revised{SN~2005el, or indeed most
of the \citet{scalzo14a} bolometric light curves.}
Moreover, the brighter lines of sight tend to have narrower bolometric light
curves, whereas \citet{scalzo14a} find the opposite trend, indicating that
asymmetry is unlikely to be the sole source of diversity in bolometric
light curves.

\citet{pakmor13} suggested
that the violent merger scenario could be extended to include mergers
of a carbon-oxygen white dwarf with a helium white dwarf, making detonations
easier to achieve and lowering the overall ejected mass.  One important
property that determines whether a double-degenerate system will merge is the
mass ratio:  more similar masses are likely to merge, less similar masses are
more likely to undergo (non-dynamical) stable Roche lobe overflow.
Since helium white dwarf masses lie in a narrower mass range and are less
massive compared to carbon-oxygen white dwarfs, the systems that do merge
will occur in the low-mass end of the primary's mass distribution, so their
masses/densities will be too low to satisfy the criteria needed for a SN~Ia
explosion.  Systems with carbon-oxygen WD masses that \emph{are} large enough
to synthesize enough \nickel\ in the explosion tend to be found among
non-merging populations
\citep[those with stable mass transfer; see Figure 2 of][]{ruiter14}.
However, the conditions typically assumed to lead to a merger in binary
population synthesis codes may be too conservative \citep[e.g.,][]{toonen14}.
Merger scenarios between carbon-oxygen and helium-rich WDs indeed warrants
further exploration \citep[e.g.][]{dan12}, since such scenarios may
further expand the allowed parameter space for violent mergers, or pick out
particular regions of interest.

In principle, variations in the fraction of the secondary's mass ejected,
or in the amount of \nickel\ produced in a dynamical merger
\citep{zhu13,dan14}, allow violent mergers to occupy the entire \Mej-\MNi\
plane above the \citet{fink10} double-detonation boundary shown in
Figure~\ref{fig:mejvmni}.  The challenge for violent merger
models is then to explain why more SNe~Ia are \emph{not} observed far from
the locus of points corresponding to the familiar width-luminosity relation.


\subsection{White dwarf collisions}

Collisions of two white dwarfs \citep{benz89} have been suggested as a means
to produce SNe~Ia.  Except in very dense environments such as globular
clusters, the rate of white dwarf collisions is expected to be very low.
However, Kozai resonances in triple systems can decrease the delay time to
a collision to increase the rate of SNe~Ia from these systems
\citep{thompson11,kd12,hamers13}.  Simulations have shown that
explosion will occur following collision \citep{rosswog09,raskin10,kushnir13},
and calculations of light curves and spectra suggest that reasonable agreement
with observations of normal SNe~Ia may be possible \citep{rosswog09}.
Furthermore, \citet{dong14} present a sample of nebular spectra of SNe~Ia
that show double-peaked line profiles, which they interpret as evidence for
axisymmetric (but aspherical) explosions characteristic of white dwarf
collisions.

This scenario shares many of the attractive qualities of
the violent merger scenario:  compared to the double-detonation scenario,
the collision model predicts higher \Mej\ (for given \MNi),
and can accomodate a significant range of ejected mass by allowing for
variation in the mass ratio of the colliding pair \citep{kushnir13}.
However, due to large uncertainties in the distribution of properties of
triple systems, this scenario currently makes no definite prediction for the
expected absolute rate of collision events, for the distribution of
masses of the colliding white dwarfs, or for the aggregate delay time
distribution of white dwarf collisions.  The only predictions currently made
for these models are positions in the \Mej-\MNi\ plane.

We see in Figure~\ref{fig:mejvmni} that the models of \citet{kushnir13}
lie close to some of our slower-declining SNe~Ia.  Interestingly, the
cosmological width-luminosity relation intersects the \Mej-\MNi\ relation
for equal-mass white dwarf collisions near $\Mej = 1.4~\Msol$.  However, none
of the current \citet{kushnir13} collision models appear in the correct
region of the \Mej-\MNi\ plane to reproduce the fastest-declining normal
SNe~Ia.  There are also several merger models on the high-mass,
slow-declining end that are not realized in the observations.

As in the discussion of double detonations, a calibration error of 0.2~\Msol\
in our \Mej\ estimates could in principle shift the entire \Mej-\MNi\
distribution up by 0.2~\Msol, which would improve the correspondence between
the full range of SNe~Ia we observe and a subset of the \citet{kushnir13}
unequal-mass white dwarf collisions.  There is no particular motivation to do
this on naturalness grounds, however:  without predictions for relative rates
of collisions between white dwarfs of different masses and mass ratios, we
have no compelling reason to prefer any subset of these models to any other
subset.


\section{Conclusions}

We have estimated \nickel\ masses and ejected masses for a large sample of
normal SNe~Ia, carefully selected to be unbiased with respect to their host
galaxy environments and explosion parameters.  The estimates are based on
empirical relations derived from the more detailed bolometric light curve
modeling of \citet{scalzo14a}.  Applying hierarchical Bayesian inference to
this sample, we have derived the intrinsic joint distribution between
\nickel\ mass and ejected mass in SNe~Ia, finding that it closely follows
the empirical one-parameter SN~Ia width-luminosity relation.  At least
one-quarter of normal SNe~Ia are sub-Chandrasekhar-mass at high confidence,
and the rate may be as high as one-half.  The fraction of normal SNe~Ia with
super-Chandrasekhar-mass ejecta appears to be quite small, about 1\%.

While our distribution is not obviously bimodal, we find that its properties
cannot be adequately described by any single explosion model.  The main
constraints can be summarized as follows:
\begin{enumerate}
\item \emph{Chandrasekhar-mass models:}  These explain many SNe~Ia well.
      However, to match bolometric light curves in the fastest-declining
      quartile of normal SNe~Ia, Chandrasekhar-mass models must show a
      narrow bolometric light curve typical of short diffusion times,
      while emitting substantially more radiation around maximum light
      than generated by radioactivity.  These features do not seem to be
      supported by contemporary Chandrasekhar-mass explosion models.
\item \emph{Double detonations:}  These explain some fast-declining SNe~Ia
      well, but (as pure detonations) produce too much \nickel\ to explain
      normal SNe~Ia outside of a narrow range of ejected masses.
\item \emph{Violent white dwarf mergers:}  These have the potential to
      explain a wide range of \nickel\ masses and ejected masses, depending
      on how efficiently they synthesize \nickel\ and how much mass they
      eject.  It remains to be seen whether the width-luminosity relation
      can be explained as a consequence of the details of the merger dynamics
      or population synthesis; more research is needed to make definite
      predictions.
\item \emph{White dwarf collisions:}  The best existing models fit observed
      Chandrasekhar-mass explosions well, but not most sub-Chandrasekhar-mass
      explosions, unless the calibration of our method is quite wrong.
      As with violent mergers, our observed distribution could place
      constraints on which actual systems can collide to form SNe~Ia.
\end{enumerate}
As a caveat, white dwarf mergers and collisions are expected to be
highly aspherical.  This explicitly breaks the approximation of spherical
symmetry used in \citet{scalzo14a}, and may give rise to line-of-sight
effects that cause the true values of \Mej\ and \MNi\ to deviate from our
estimates to some extent.
However, these effects should tend to increase diversity rather than
decreasing it, and there are observational reasons (e.g., polarization) to
believe that asymmetries are modest rather than extreme.  There is thus no
compelling reason to believe that asymmetry plays a major role, unless it
drives a low-scatter width-luminosity relation within the context of a
specific scenario.

In fact, the full \Mej-\MNi\ distribution is currently beyond the capacity of
any single explosion model or progenitor scenario to predict, owing partly
to challenges in modeling explosions and partly to challenges in modeling
stellar populations.  As theoretical predictions improve, our distribution
will provide strong constraints on the entire end-to-end physical
processes resulting in SNe~Ia, from initial formation of the system through
to the final explosion.  Our methodology is simple, relying only on photometry
that will be collected for any SN~Ia cosmology experiment, and can be
repeated in the future with much larger data sets, such as those provided by
the Dark Energy Survey (DES) and Large Synoptic Survey Telescope (LSST).
With enough events from untargeted surveys at both low and high redshifts,
the delay time $t_\mathrm{delay}$ can also be folded in, enabling comparison
of end-to-end modeling to the full \emph{trivariate}
\Mej-\MNi-$t_\mathrm{delay}$ distribution.  Such complete information about
the distributions of progenitor properties could allow us not only to
determine, authoritatively, which progenitor and explosion channels contribute
to SNe~Ia, but in what proportions and in what regions of parameter space,
enabling new breakthroughs both in cosmology and in stellar evolution.


\section*{Acknowledgments}

Parts of this research were conducted by the Australian Research Council
Centre of Excellence for All-Sky Astrophysics (CAASTRO), through project
number CE110001020.
RS and AJR acknowledge support from ARC Laureate Grant FL0992131.
We thank Ivo Seitenzahl and Brian Schmidt for useful discussions.



\begin{thebibliography}{199}

\footnotesize

\bibitem[\protect\citeauthoryear{Arnett}{1982}]{arnett82}
Arnett, W. D.
   1982, \apj, 253, 785

\bibitem[\protect\citeauthoryear{Anand}{1965}]{anand65}
Anand, S. P. S.
   1965, Proc. Natl. Acad. Sci., 54, 23

\bibitem[\protect\citeauthoryear{Astier \etal}{2006}]{snls}
Astier, P., Guy, J., Regnault, N., \etal\
   2006, \apj, 447, 31

\bibitem[\protect\citeauthoryear{Bailey \etal}{2009}]{bailey09}
Bailey, S., Aldering, G., Antilogus, P., \etal\
   2009, \aanda, 500, L17

\bibitem[\protect\citeauthoryear{Baron \etal}{2012}]{baron12}
Baron, E., H\"oflich, P., Krisciunas, K., \etal\
   2012, \apj, 753, 105

\bibitem[\protect\citeauthoryear{Betoule \etal}{2014}]{betoule14}
Betoule, M., Kessler, R., Guy, J., \etal\
   2014, \aanda, submitted (\arxiv{1401:4064})

\bibitem[\protect\citeauthoryear{Benz \etal}{1989}]{benz89}
Benz, W., Thielemann, F.-K., \& Hills, J.~G.
   1989, \apj, 342, 986

\bibitem[\protect\citeauthoryear{Blinnikov \& Sorokina }{2010}]{bs10}
Blinnikov, S.~I. \& Sorokina, E.~I.
   2010, \arxiv{1009.4353}

\bibitem[\protect\citeauthoryear{Bloom \etal}{2012}]{bloom12}
Bloom, J.~S., Kasen, D., Shen, K.~J., \etal\
   2012, \apj, 744, L17

\bibitem[\protect\citeauthoryear{Blondin \etal}{2012}]{blondin12}
Blondin, S., Matheson, T., Kirshner, R.~P., \etal\
   2012, \aj, 143, 5

\bibitem[\protect\citeauthoryear{Blondin \etal}{2013}]{blondin13}
Blondin, S., Dessart, L., Hillier, D.~J., \etal\
   2013, \mnras, 429, 2127

\bibitem[\protect\citeauthoryear{Bongard \etal}{2006}]{bongard06}
Bongard, S., Baron, E., Smadja, G., \etal\
   2006, \apj, 647, 513

\bibitem[\protect\citeauthoryear{Branch \etal}{1993}]{bfn93}
Branch, D., Fisher, A., \& Nugent, P.
   1993, \aj, 106, 2383

\bibitem[\protect\citeauthoryear{Branch \etal}{2006}]{branch06}
Branch, D., Dang, L., Hall, N., \etal\
   2006, \pasp, 118, 560

\bibitem[\protect\citeauthoryear{Branch \etal}{2007}]{branch07}
Branch, D., Troxel, M.~A., Jeffery, D.~J., \etal\
   2007, \pasp, 119, 135

\bibitem[\protect\citeauthoryear{Branch \etal}{2008}]{branch08}
Branch, D., Jeffery, D.~J., Parrent, J., \etal\
   2008, \pasp, 120, 135

\bibitem[\protect\citeauthoryear{Burns \etal}{2014}]{burns14}
Burns, C., Stritzinger, M., Phillips, M.~M., \etal\
   2014, \apj, in press (\arxiv{1405.3934})

\bibitem[\protect\citeauthoryear{Cardelli \etal}{1988}]{cardelli}
Cardelli, J.~A., Clayton, G.~C. \& Mathis, J.~S.
   1988, \apj, 329, L33

\bibitem[\protect\citeauthoryear{Chamulak \etal}{2008}]{chamulak08}
Chamulak, D.~A., Brown, E.~F., Timmes, F.~X., \etal\
   2008, \apj, 677, 160

\bibitem[\protect\citeauthoryear{Chotard \etal}{2011}]{chotard11}
Chotard, N., Gangler, E., Aldering, G., \etal\
   2011, \aanda, 529, 4

\bibitem[\protect\citeauthoryear{Churazov \etal}{2014}]{integral}
Churazov, E., Sunyaev, R., Isern, J., \etal\
   2014, \arxiv{1405.3332}

\bibitem[\protect\citeauthoryear{Conley \etal}{2006}]{conley06}
Conley, A., Goldhaber, G., Wang, L., \etal\
   2006, \apj, 644, 1

\bibitem[\protect\citeauthoryear{Dan \etal}{2012}]{dan12}
Dan, M., Rosswog, S., Guillochon, J., \etal\
   2012, \mnras, 422, 2417

\bibitem[\protect\citeauthoryear{Dan \etal}{2013}]{dan14}
Dan, M., Rosswog, S., Br\"uggen, M., \etal\
   2014, \mnras, 438, 14

\bibitem[\protect\citeauthoryear{Di~Stefano}{2010a}]{rds10a}
Di~Stefano, R.
   2010, \apj, 712, 728

\bibitem[\protect\citeauthoryear{Di~Stefano}{2010b}]{rds10b}
Di~Stefano, R.
   2010, \apj, 719, 474

\bibitem[\protect\citeauthoryear{Di~Stefano \& Kilic}{2012}]{rds12}
Di~Stefano, R. \& Kilic, M.
   2012, \apj, 759, 56

\bibitem[\protect\citeauthoryear{Dong \etal}{2014}]{dong14}
Dong, S., Katz, B., Kushnir, D., \etal\
   2014, \arxiv{1401.3347}

\bibitem[\protect\citeauthoryear{Fink \etal}{2010}]{fink10}
Fink, M., R\"opke, F.~K., Hillebrandt, W., \etal\
   2010, \aanda, 514, A53

\bibitem[\protect\citeauthoryear{Filippenko \etal}{1992}]{filippenko92}
Filippenko, A.~V., Richmond, M.~W., Matheson, T., \etal\
   1992, \apj, 384, L15

\bibitem[\protect\citeauthoryear{Fink \etal}{2014}]{fink14}
Fink, M., Kromer, M., Seitenzahl, I.~R., \etal\
   2014, \mnras, 438, 1762

\bibitem[\protect\citeauthoryear{Folatelli \etal}{2010}]{csp10}
Folatelli, G., Phillips, M.~M., Burns, C., R., \etal\
   2010, \aj, 139, 120

\bibitem[\protect\citeauthoryear{Foley \& Kasen}{2011}]{fk11}
Foley, R.~J. \& Kasen, D.
   2010, \apj, 729, 55

\bibitem[\protect\citeauthoryear{Foreman-Mackey \etal}{2013}]{emcee}
Foreman-Mackey, D., Hogg, D.~W., Lang, D., \etal\
   2013, \pasp, 125, 306

\bibitem[\protect\citeauthoryear{Frieman \etal}{2008}]{sdss}
Frieman, J.~A., Bassett, B., Becker, A., \etal\
   2008, \aj, 135, 338

\bibitem[\protect\citeauthoryear{Fryer \etal}{2010}]{fryer10}
Fryer, C.~L., Ruiter, A.~J., Belczynski, K., \etal\
   2010, ApJ, 725, 296

\bibitem[\protect\citeauthoryear{Goldhaber \etal}{2001}]{goldhaber01}
Goldhaber, G., Groom, D.~E., Kim, A.~G., \etal\
   2001, \apj, 558, 359

\bibitem[\protect\citeauthoryear{Gilfanov \& Bogdan}{2010}]{gb10}
Gilfanov, M. \& Bogd\'an, \'A.
   2010, Nature, 463, 924

\bibitem[\protect\citeauthoryear{Guy \etal}{2007}]{guy07}
Guy, J., Astier, P., Baumont, S., \etal\
   2007, \aanda, 466, 11

\bibitem[\protect\citeauthoryear{Guy \etal}{2010}]{guy10}
Guy, J., Sullivan, M., Conley, A., \etal\
   2010, \aanda, 523, 7

\bibitem[\protect\citeauthoryear{Hachisu \etal}{2011}]{hachisu11}
Hachisu, I., Kato, M., Saio, H., \etal\
   2012, \apj, 744, 69

\bibitem[\protect\citeauthoryear{Hamers \etal}{2013}]{hamers13}
Hamers, A.~S., Pols, O.~R., Claeys, J.~S.~W., \etal\
   2013, \mnras, 430, 2262

\bibitem[\protect\citeauthoryear{Hicken \etal}{2007}]{hicken07}
Hicken, M., Garnavich, P.~M., Prieto, J.~L., \etal\
   2007, \apj, 669, L17

\bibitem[\protect\citeauthoryear{Hillebrandt \etal}{2013}]{hillebrandt13}
Hillebrandt, W., Kromer, M., R\"opke, F.~K., \& Ruiter, A.~J.
   2013, Frontiers of Physics, 8, 116

\bibitem[\protect\citeauthoryear{H\"oflich \& Khohklov}{1996}]{hk96}
H\"oflich, P. \& Khohklov, A.
   1996, \apj, 457, 500

\bibitem[\protect\citeauthoryear{H\"oflich \etal}{2004}]{hoflich04}
H\"oflich, P., Gerardy, C., Nomoto, K., \etal\
   2004, \apj, 617, 1258

\bibitem[\protect\citeauthoryear{Hogg \etal}{2010}]{hogg10}
Hogg, D., Myers, A., \& Bovy, J.
   2010, \apj, 725, 2166

\bibitem[\protect\citeauthoryear{Howell \etal}{2006}]{howell06}
Howell, D.~A., Sullivan, M., Nugent, P.~E., \etal\
   2006, Nature, 443, 308

\bibitem[\protect\citeauthoryear{Howell \etal}{2007}]{howell07}
Howell, D. A. \etal\
   2007, \apj, 667, 37

\bibitem[\protect\citeauthoryear{Howell \etal}{2009}]{howell09}
Howell, D. A. \etal\
   2009, \apj, 691, 661

\bibitem[\protect\citeauthoryear{Iben \& Tutukov}{1984}]{it84}
Iben, I. \& Tutukov, A.~V.
   1984, \apjs, 54, 335 

\bibitem[\protect\citeauthoryear{Iben \& Tutukov}{1991}]{it91}
Iben, I. \& Tutukov, A.~V.
   1991, \apj, 370, 615

\bibitem[\protect\citeauthoryear{Jha \etal}{2006}]{jha06}
Jha, S., Branch, D., Chornock, R., \etal\
   2006, \aj, 132, 189

\bibitem[\protect\citeauthoryear{Jordan \etal}{2012}]{jordan12}
Jordan, G.~C., Perets, H.~B., Fisher, R.~T., \etal\
   2012, \apj, 761, L23

\bibitem[\protect\citeauthoryear{Justham}{2011}]{justham11}
Justham, S.
   2011, \apj, 730, L34

\bibitem[\protect\citeauthoryear{Kasen \etal}{2009}]{kasen09}
Kasen, D., R\"opke, F.~K., \& Woosley, S.~E.
   2009, Nature, 460, 869

\bibitem[\protect\citeauthoryear{Katz \& Dong}{2012}]{kd12}
Katz, B. \& Dong, S.
   2012, \arxiv{1211.4584}

\bibitem[\protect\citeauthoryear{Khokhlov}{1991}]{khokhlov91}
Khokhlov, A.
   1991, \aanda, 245, 114

\bibitem[\protect\citeauthoryear{Khokhlov \etal}{1993}]{kmh93}
Khokhlov, A., M\"uller, E. \& H\"oflich, P.
   1993, \aanda, 270, 223

\bibitem[\protect\citeauthoryear{Krisciunas \etal}{2011}]{krisciunas11}
Krisciunas, K., Li, W., Matheson, T., \etal\
   2011, \aj, 142, 74

\bibitem[\protect\citeauthoryear{Kromer \etal}{2013a}]{kromer13a}
Kromer, M., Fink, M., Stanishev, V., \etal\
   2013, \mnras, 429, 2287

\bibitem[\protect\citeauthoryear{Kromer \etal}{2013b}]{kromer13b}
Kromer, M., Pakmor, R., Taubenberger, S., \etal\
   2013, \apj, 778, 18

\bibitem[\protect\citeauthoryear{Krueger \etal}{2012}]{krueger12}
Krueger, B.~K. Jackson, A.~P., Townsley, D.~M., \etal
   2012, \apj, 757, 175

\bibitem[\protect\citeauthoryear{Kushnir \etal}{2013}]{kushnir13}
Kushnir, D., Katz, B., Dong, S., \etal\
   2013, \apj, 778, L37

\bibitem[\protect\citeauthoryear{Li \etal}{2011}]{li11}
Li, W., Leaman, J.; Chornock, R., \etal
   2011, \mnras, 412, 1441

\bibitem[\protect\citeauthoryear{Livne \etal}{2005}]{livne05}
Livne, E., Asida, S.~M., \& H\"oflich, P.
   2005, \apj, 632, 443

\bibitem[\protect\citeauthoryear{Lesaffre \etal}{2006}]{lesaffre06}
Lesaffre, P., Han, Z., Tout, C.~A., \etal
   2006, \mnras, 368, 187

\bibitem[\protect\citeauthoryear{Long \etal}{2013}]{long13}
Long, M., Jordan, G.~C., van Rossum, D.~R., \etal\
   2013, \arxiv{1307.8221}

\bibitem[\protect\citeauthoryear{Maeda \etal}{2011}]{maeda11}
Maeda, K., Leloudas, S., Taubenberger, S., \etal\
   2011, \mnras, 413, 3075


\bibitem[\protect\citeauthoryear{Mandel \etal}{2009}]{mandel09}
Mandel, K.~S., Wood-Vasey, W.~M., Friedman, A.~S., \etal\
   2009, \apj, 704, 629

\bibitem[\protect\citeauthoryear{Mandel \etal}{2011}]{mandel11}
Mandel, K.~S., Narayan, G., \& Kirshner, R.~P.
   2011, \apj, 731, 120

\bibitem[\protect\citeauthoryear{Mannucci \etal}{2006}]{mannucci06}
Mannucci, F., Della Valle, M., Panagia, N.
   2006, \mnras, 370, 773

\bibitem[\protect\citeauthoryear{Mazzali \etal}{1998}]{mazzali98}
Mazzali, P.~A., Cappellaro, E., Danziger, I.~J., \etal\
   1998, \apj, 499, L49

\bibitem[\protect\citeauthoryear{Mazzali \etal}{2007}]{mazzali07}
Mazzali, P. A., R\"opke, F., Benetti, S., \etal\
   2007, Science, 315, 825

\bibitem[\protect\citeauthoryear{Mazzali \etal}{2011}]{mazzali11}
Mazzali, P.~A., Maurer, I., Stritzinger, M., \etal\
   2011, \mnras, 416, 881

\bibitem[\protect\citeauthoryear{Mazzali \& Hachinger}{2012}]{mh12}
Mazzali, P.~A., \& Hachinger, S.
   2012, \mnras, 424, 2926

\bibitem[\protect\citeauthoryear{Moll \etal}{2014}]{moll14}
Moll, R., Raskin, C., Kasen, D., \etal\
   2014, \apj, 785, 105

\bibitem[\protect\citeauthoryear{Motohara \etal}{2006}]{motohara06}
Motohara, K., Maeda, K., Gerardy, C.~L., \etal\
   2006, \apj, 652, L101


\bibitem[\protect\citeauthoryear{Nomoto \& Kondo}{1991}]{nk91}
Nomoto, K. \& Kondo, Y.
   1991, \apj, 367, L19

\bibitem[\protect\citeauthoryear{Nugent \etal}{1995}]{nugent95}
Nugent, P., Branch, D., Baron, E., \etal\
   1995, \prl, 75, 394

\bibitem[\protect\citeauthoryear{Nugent \etal}{2011}]{nugent11}
Nugent, P., Sullivan, M., Cenko, S. B., \etal\
   2011, Nature, 480, 344


\bibitem[\protect\citeauthoryear{Pakmor \etal}{2010}]{pakmor10}
Pakmor, R., Kromer, M., R\"opke, F.~K., \etal
   2010, Nature, 463, 61

\bibitem[\protect\citeauthoryear{Pakmor \etal}{2011}]{pakmor11}
Pakmor, R., Hachinger, S., R\"opke, F.~K., \etal
   2011, \aanda, 528, 117

\bibitem[\protect\citeauthoryear{Pakmor \etal}{2012}]{pakmor12}
Pakmor, R., Kromer, M., \& Taubenberger, S.
   2013, \apj, 747, L10

\bibitem[\protect\citeauthoryear{Pakmor \etal}{2013}]{pakmor13}
Pakmor, R., Kromer, M., Taubenberger, S., \etal
   2013, \apj, 770, L8

\bibitem[\protect\citeauthoryear{Perlmutter \etal}{1999}]{scp99}
Perlmutter, S., Aldering, G., Goldhaber, G., \etal\
   1999, \apj, 517, 565

\bibitem[\protect\citeauthoryear{Phillips \etal}{1992}]{phillips92}
Phillips, M.~M., Wells, L.~A., Suntzeff, N.~B., \etal\
   1992, \aj, 103, 1632

\bibitem[\protect\citeauthoryear{Phillips }{1993}]{phillips93}
Phillips, M.~M.
   1993, \apj, 413, 105

\bibitem[\protect\citeauthoryear{Phillips \etal}{1999}]{phillips99}
Phillips, M.~M., Lira, P., Suntzeff, N.~B., \etal\
   1999, \aj, 118, 1766

\bibitem[\protect\citeauthoryear{Phillips \etal}{2007}]{phillips07}
Phillips, M.~M., Li, W., Frieman, J.~A., \etal\
   2007, \pasp, 119, 360

\bibitem[\protect\citeauthoryear{Pinto \& Eastman}{2000}]{pe00}
Pinto, P. \& Eastman, R.
   2000, \apj, 530, 744

\bibitem[\protect\citeauthoryear{Piro \etal}{2008}]{piro08}
Piro, A.
   2008, \apj, 679, 616

\bibitem[\protect\citeauthoryear{Piro \etal}{2010}]{piro10}
Piro, A., Chang, P., \& Weinberg, N.~N.
   2010, \apj, 708, 598

\bibitem[\protect\citeauthoryear{Piro \etal}{2014}]{piro14}
Piro, A., Thompson, T.~A., \& Kochanek, C.~S.
   2014, \mnras, 438, 3456


\bibitem[\protect\citeauthoryear{Raskin \etal}{2009}]{raskin09}
Raskin, C., Timmes, F.~X., Scannapieco, E., \etal\
   2009, \mnras, 399, L156

\bibitem[\protect\citeauthoryear{Raskin \etal}{2010}]{raskin10}
Raskin, C., Scannapieco, E., Rockefeller, G., \etal\
   2010, \apj, 724, 111

\bibitem[\protect\citeauthoryear{Riess \etal}{1996}]{riess96}
Riess, A.~G., Press, W.~H., \& Kirshner, R.~P.
   1996, \apj, 473, 88

\bibitem[\protect\citeauthoryear{Riess \etal}{1998}]{riess98}
Riess, A.~G., Filippenko, A.~V., Challis, P., \etal\
   1998, \aj, 116, 1009

\bibitem[\protect\citeauthoryear{R\"opke \etal}{2012}]{roepke12}
R\"opke, F.~K., Kromer, M., Seitenzahl, I.~R., \etal\
   2012, \apj, 750, 19

\bibitem[\protect\citeauthoryear{Rosswog \etal}{2009}]{rosswog09}
Rosswog, S., Kasen, D., Guillochon, J., \etal\
   2009, \apj, 705, 128

\bibitem[\protect\citeauthoryear{Roxburgh}{1965}]{roxburgh65}
Roxburgh, I.~W.
   1965, Z. Astrophys., 62, 134

\bibitem[\protect\citeauthoryear{Ruiter \etal}{2013}]{ruiter13}
Ruiter, A.~J., Sim, S.~A., Pakmor, R., \etal\
   2013, \mnras, 429, 1425

\bibitem[\protect\citeauthoryear{Ruiter \etal}{2014}]{ruiter14}
Ruiter, A.~J., Belczynski, K., Sim, S.~A., \etal\
   2014, \mnras, 440, L101

\bibitem[\protect\citeauthoryear{Ruiz-Lapuente}{2014}]{pilar14}
Ruiz-Lapuente, P.
   2014, \newar, submitted (\arxiv{1403.4087})

\bibitem[\protect\citeauthoryear{Sahu \etal}{2008}]{sahu08}
Sahu, D.~K., Tanaka, M., Anupama, G.~C., \etal
   2008, \apj, 680, 580

\bibitem[\protect\citeauthoryear{Scannapieco \etal}{2005}]{sb05}
Scannapieco, E., \& Bildsten, L.
   2005, \apj, 629, 85

\bibitem[\protect\citeauthoryear{Scalzo \etal}{2010}]{scalzo10}
Scalzo, R.~A., Aldering, G., Antilogus, P., \etal\
   2010, \apj, 713, 1073

\bibitem[\protect\citeauthoryear{Scalzo \etal}{2012}]{scalzo12}
Scalzo, R.~A., Aldering, G., Antilogus, P., \etal\
   2012, \apj, 757, 12

\bibitem[\protect\citeauthoryear{Scalzo \etal}{2014a}]{scalzo14a}
Scalzo, R.~A., Aldering, G., Antilogus, P., \etal\
   2014, \mnras, 440,1498

\bibitem[\protect\citeauthoryear{Scalzo \etal}{2014b}]{scalzo14b}
Scalzo, R.~A., Childress, M., Tucker, B., \etal\
   2014, \mnras, submitted (\arxiv{1404.1002})

\bibitem[\protect\citeauthoryear{Scolnic \etal}{2013}]{scolnic13}
Scolnic, D., Riess, A., Foley, R.~J., \etal\
   2014, \apj, 780, 1


\bibitem[\protect\citeauthoryear{Seitenzahl \etal}{2011}]{seitenzahl11}
Seitenzahl, I., Ciaraldi-Schoolmann, F., \& R\"opke, F.~K.
   2011, \mnras, 414, 2709

\bibitem[\protect\citeauthoryear{Seitenzahl \etal}{2013a}]{seitenzahl13a}
Seitenzahl, I., Ciaraldi-Schoolmann, F., R\"opke, F.~K., \etal\
   2013, \mnras, 429, 1156

\bibitem[\protect\citeauthoryear{Seitenzahl \etal}{2013b}]{seitenzahl13b}
Seitenzahl, I., Cescutti, G., R\"opke, F.~K., \etal
   2013, \aanda, 559, L5

\bibitem[\protect\citeauthoryear{Sim \etal}{2010}]{sim10}
Sim, S.~A., R\"opke, F.~K., Hillebrandt, W., \etal\
   2010, \apj, 714, L52

\bibitem[\protect\citeauthoryear{Sim \etal}{2013}]{sim13}
Sim, S.~A., Seitenzahl, I.~R., R\"opke, F.~K., \etal\
   2013, \mnras, 436, 333

\bibitem[\protect\citeauthoryear{Shen \etal}{2012}]{shen12}
Shen, K., Bildsten, L., Kasen, D., \etal\
   2012, \apj, 748, 35


\bibitem[\protect\citeauthoryear{Stehle \etal}{2005}]{stehle05}
Stehle, M., Mazzali, P.~A., Benetti, S., \etal\
   2005, \mnras, 360, 1231

\bibitem[\protect\citeauthoryear{Stritzinger \etal}{2006}]{stritz06}
Stritzinger, M., Leibundgut, B., Walch, S., \etal\
   2006, \aanda, 450, 241

\bibitem[\protect\citeauthoryear{Sullivan \etal}{2006}]{sullivan06}
Sullivan, M., Le Borgne, D., Pritchet, C.~J., \etal\
   2006, \apj, 648, 868

\bibitem[\protect\citeauthoryear{Tanaka \etal}{2010}]{tanaka10}
Tanaka, M., Kawabata, K., Yamanaka, M., \etal\
   2010, \apj, 714, 1209

\bibitem[\protect\citeauthoryear{Taubenberger \etal}{2011}]{taub11}
Taubenberger, S., Benetti, S., Childress, M. \etal\
   2011, \mnras, 412, 2735

\bibitem[\protect\citeauthoryear{Taubenberger \etal}{2013}]{taub13}
Taubenberger, S., Kromer, M., Hachinger, S. \etal\
   2013, \mnras, 432, 3117

\bibitem[\protect\citeauthoryear{Thompson \etal}{2011}]{thompson11}
Thompson, T.~A.
   2011, \apj, 741, 82

\bibitem[\protect\citeauthoryear{Toonen \etal}{2014}]{toonen14}
Toonen, S., Claeys, J.~S.~W., Mennekens, N., \etal\
   2014, \aanda, 562, 14

\bibitem[\protect\citeauthoryear{Tripp}{1998}]{tripp98}
Tripp, R.
   1998, \aanda, 331, 815

\bibitem[\protect\citeauthoryear{van Kerkwijk \etal}{2010}]{vkcj10}
van Kerkwijk, M., Chang, P., \& Justham, S.
   2010, \apj, 722, L157

\bibitem[\protect\citeauthoryear{Wang \& Han}{2012}]{wh12}
Wang, B., \& Han, Z.
   2012, New Astronomy Reviews, 56, 122

\bibitem[\protect\citeauthoryear{Wang \& Wheeler}{2008}]{ww08}
Wang, L., \& Wheeler, J.~C.
   2008, \araa, 46, 433

\bibitem[\protect\citeauthoryear{Wang \etal}{2009}]{wang09}
Wang, X., Filippenko, A.~V., Ganeshalingam, M., \etal\
   2009, \apj, 699, L139

\bibitem[\protect\citeauthoryear{Whelan \& Iben}{1973}]{wi73}
Whelan, J. \& Iben, I. J.
   1973, \apj, 186, 1007

\bibitem[\protect\citeauthoryear{Woods \& Gilfanov}{2013}]{wg13}
Woods, T.~E., \& Gilfanov, M.
   2013, \mnras, 432, 1640

\bibitem[\protect\citeauthoryear{Woosley \& Weaver}{1994}]{ww94}
Woosley, S.~E. \& Weaver, T. A.
   1994, \apj, 423, 371

\bibitem[\protect\citeauthoryear{Woosley \& Kasen}{2007}]{wk07}
Woosley, S.~E. \& Kasen, D.
   2007, \apj, 662, 487

\bibitem[\protect\citeauthoryear{Woosley \& Kasen}{2011}]{wk11}
Woosley, S.~E. \& Kasen, D.
   2011, \apj, 734, 38

\bibitem[\protect\citeauthoryear{Yamanaka \etal}{2009}]{yamanaka09}
Yamanaka, M., Kawabata, K., Kinugasa, K., \etal\
   2009, \apj, 707, L118

\bibitem[\protect\citeauthoryear{Yoon \& Langer}{2005}]{yl05}
Yoon, S.-C. \& Langer, N.
   2005, \aanda, 435, 967

\bibitem[\protect\citeauthoryear{Yuan \etal}{2010}]{yuan10}
Yuan, F., Quimby, R.~M., Wheeler, J.~C., \etal\
   2010, \apj, 715, 1338

\bibitem[\protect\citeauthoryear{Zhu \etal}{2013}]{zhu13}
Zhu, C., Chang, P., van Kerkwijk, M., \etal\
   2013, \apj, 767, 164

\end{thebibliography}
\end{document}